\documentclass{article}
\PassOptionsToPackage{table}{xcolor}
\usepackage{iclr2025_conference}

\usepackage{url}
\usepackage{xspace}
\usepackage{amssymb}
\usepackage{xcolor}
\usepackage{multirow}
\usepackage{multicol}
\usepackage{float}
\usepackage{booktabs}
\usepackage{amsmath,amsthm}
\usepackage{pifont}
\usepackage{hyperref}
\IfFileExists{accessibility.sty}{%
  \usepackage{accessibility}%
}{%
  \PackageWarningNoLine{IntTravel}{accessibility.sty unavailable; continuing without tagged PDF support}%
}

\DeclareMathOperator*{\argmax}{arg\,max}
\newcommand{\best}[1]{\cellcolor{dreamblue!15}#1}
\newcommand{\second}[1]{\textcolor{dreamblue}{#1}}

\hypersetup{
  colorlinks=true,
  linkcolor=dreamblue,
  citecolor=dreamblue,
  urlcolor=dreamblue,
  filecolor=dreamblue,
}

\renewcommand{\dreamxrunhead}{DreamX}
\renewcommand{\dreamxvenue}{DreamX}
\renewcommand{\dreamxdate}{September 2026}

\title{IntTravel: A Real-World Dataset and Generative Framework for Integrated Multi-Task Travel Recommendation}

\author{%
Huimin Yan\thanks{Equal contribution.}, Longfei Xu\footnotemark[1] \thanks{Project lead.}, Junjie Sun\footnotemark[1], Zheng Liu,
\\
Wei Luo, Kaikui Liu, Xiangxiang Chu
\AND
{\normalfont Amap, Alibaba Group}%
}

\iclrfinalcopy

\begin{document}

\maketitle

\begin{abstract}
Next Point of Interest (POI) recommendation is essential for modern mobility and location-based services. To provide a smooth user experience, models must understand several components of a journey holistically: ``when to depart'', ``how to travel'', ``where to go'', and ``what needs arise via the route''. However, current research is limited by fragmented datasets that focus merely on next POI recommendation (``where to go''), neglecting the departure time, travel mode, and situational requirements along the journey. Furthermore, the limited scale of these datasets impedes accurate evaluation of performance. To bridge this gap, we introduce \textbf{IntTravel}, the first large-scale public dataset collected from Amap for integrated travel recommendation, including \textbf{4.1 billion interactions from 163 million users with 7.3 million POIs}. Built upon this dataset, we introduce an end-to-end, decoder-only generative framework for multi-task recommendation. It incorporates information preservation, selection, and factorization to balance task collaboration with specialized differentiation, yielding substantial performance gains. IntTravel has been successfully deployed on Amap serving hundreds of millions of users, leading to a 1.09\% increase in CTR. IntTravel is available at \url{https://github.com/AMAP-ML/DreamX-Rec/}.
\end{abstract}

\section{Introduction}

Next POI~(Point of Interest) prediction is crucial for location-based service platforms. Accurate next-POI prediction enables high-quality services for hundreds of millions of users. Beyond next-POI prediction, a truly seamless service requires modeling a user's intents across multiple dimensions: when to depart, where to go, how to travel, and what needs arise via the route.

However, we find that existing datasets~\citep{cho2011friendship,cheng2011exploring,yang2014modeling,yang2016participatory,monti2018semantic,yang2019revisiting} in travel recommendation face two significant limitations~(see Table~\ref{Table_dataset_comparison}). \textbf{Most datasets are generated and processed for next-POI recommendation}, while ignoring the relationships among destination, departure time, travel mode, and users' on-the-way intentions. Furthermore, \textbf{the insufficient scale of these datasets hinders reliable evaluations of generative models}, potentially leading to inaccurate findings. To bridge these gaps, we introduce \textbf{IntTravel}, a comprehensive dataset for multi-task travel recommendation. It is built from large-scale, anonymized industrial data collected from Amap, one of the world's largest mobility and location-based service platforms. IntTravel covers four core tasks that jointly characterize a user's journey:

\begin{itemize}
    \item \textbf{When (departure time)}: Predicting the time a user intends to start a journey.
    \item \textbf{Where (destination)}: Recommending the POI the user aims to reach in their next journey.
    \item \textbf{How (travel mode)}: Determining the preferred mode, such as driving, walking, or public transit.
    \item \textbf{Via (on-the-way intention)}: Predicting the via POI a user is likely to visit while traveling toward a known destination.
\end{itemize}

\begin{table*}[!t]
  \centering
  \caption{Comparison of travel recommendation datasets.}
  \label{Table_dataset_comparison}
  \resizebox{\textwidth}{!}{%
  \begin{tabular}{lrrcccc}
    \toprule
    \multirow{2}{*}{\textbf{Dataset}} & \multirow{2}{*}{\textbf{\# Users}} & \multirow{2}{*}{\textbf{\# Interactions}} & \multicolumn{4}{c}{\textbf{Data}} \\
    \cmidrule(lr){4-7}
    & & & POI check-in & Time feature & Travel mode & Via POIs \\
    \midrule
    Cheng et al. (2011) & 224,804 & 22,388,315 & \ding{51} & \ding{51} & \ding{55} & \ding{55} \\
    Cho et al. (2011) & 196,591 & 6,442,890 & \ding{51} & \ding{51} & \ding{55} & \ding{55} \\
    Yang et al. (2014) & 2,763 & 801,131 & \ding{51} & \ding{51} & \ding{55} & \ding{55} \\
    Yang et al. (2016) & 266,909 & 33,278,683 & \ding{51} & \ding{51} & \ding{55} & \ding{55} \\
    Monti et al. (2018) & 424,730 & 12,473,360 & \ding{51} & \ding{51} & \ding{55} & \ding{55} \\
    Yang et al. (2019) & 114,324 & 22,809,624 & \ding{51} & \ding{51} & \ding{55} & \ding{55} \\
    \midrule
    \textbf{IntTravel (ours)} & \textbf{163 M} & \textbf{4.1 B} & \ding{51} & \ding{51} & \ding{51} & \ding{51} \\
    \bottomrule
  \end{tabular}
  }
\end{table*}

Recently, models for next-POI prediction have transitioned from traditional deep learning architectures to generative paradigms~\citep{li2024large,wang2024large,wang2025generative,wei2025oneloc}. By formulating recommendation as a sequence generation problem, generative models can capture complex dependencies and reduce reliance on complex feature engineering~\citep{Tiger_rajput2023,HSTU_zhai2024,yan2025intsr,OneRec_deng2025}. These models achieve comparable or even better performance than traditional deep-learning-based models.

Nevertheless, traditional multi-task frameworks are difficult to apply directly to generative architectures. This is mainly because \textbf{sequence representation has evolved from an important feature into the primary driver for next-token generation}. Traditional methods process tasks after combining information, which means they fail to maintain task-specific independence within the Transformer's encoding layers. To tackle this issue, we introduce \textbf{IntTravel}, the first multi-task generative framework for end-to-end travel recommendation. The architecture is built upon three primary components: (1) \textbf{Task-Guided Information Persistence}: This module enriches hyper-connections with task-specific properties, thereby streamlining the delivery of information across the architecture and ensuring its integrity. (2) \textbf{Task-Specific Selective Gating}: This module employs selective gating across hidden layers to retrieve multi-scale, task-specific information tailored to different encoding levels. (3) \textbf{Task-Aware Scenario Factorization}: This module combines the outputs of the persistence and selective stages. It uses shared and separate parameters among experts to balance common knowledge with specific task needs.

Comprehensive offline experiments demonstrate that IntTravel consistently outperforms all compared deep-learning-based multi-task baselines on the IntTravel dataset. The framework has been fully deployed within Amap, powering personalized POI recommendation services. In online A/B tests, IntTravel achieved a significant 1.09\% relative increase in POI Click-Through Rate (CTR) over a strong, non-multi-task generative baseline. 

In summary, our contributions are as follows:

\begin{itemize}
    \item \textbf{Real-world industrial dataset}: We introduce IntTravel, a large-scale, real-world dataset that integrates four key travel tasks (When, How, Where, and Via).
    \item \textbf{Generative multi-task framework}: We design IntTravel, a multi-task decoder-only framework that integrates four travel tasks as a joint sequence generation problem, enabling end-to-end optimization.
    \item \textbf{Offline demonstrations and online deployment}: We validate IntTravel through comprehensive offline evaluation and online A/B testing. We detail the design, evaluation, and deployment of IntTravel at Amap, offering insights into its real-world effectiveness and scalability.
\end{itemize}

\section{IntTravel Dataset}
\label{Section_inttravel_dataset}

IntTravel\footnote{\url{https://huggingface.co/datasets/GD-ML/IntTravel_dataset}} is a comprehensive real-world dataset for multi-task travel recommendation, including 162,815,693 users, 7,291,872 POIs, and 4,129,827,011 user interactions (mean: 25, median: 21). It captures four interconnected dimensions of a user's journey: departure time, travel mode, destination, and via-POI intent. Detailed data fields, distributions, and examples are provided in Appendix~\ref{Appendix_data_statistics_and_examples}.

\subsection{Data Collection}
All data were collected from Amap with the explicit consent of each user. The dataset comprises anonymized user profiles, POIs and their associated information, and raw user operational logs. These logs span a 90-day period across 10 major cities in China and record the following types of user behavior: (1) search and click actions on POIs within the application; (2) ordering actions for a specific POI, such as purchasing a service or dining coupon; (3) route planning, including the selected mode of travel; (4) navigation initiation, including the selected mode of travel; and (5) searching for and adding via POIs along a route.

Because travel behavior is closely associated with spatiotemporal context, the logs also include the time, geographic block ID, and weather conditions for each action.

\subsection{User Profile Anonymization}
To protect user privacy, we applied an anonymization process to the raw data. Original user IDs were randomly shuffled and re-indexed using consecutive integer IDs. The original names of the six profile features in the released dataset were concealed and represented by generic labels such as ``profile feature 1'' and ``profile feature 2.'' Values for each feature were mapped to discrete, consecutive integer IDs, preventing users of the dataset from inferring their real-world meanings. Raw latitude and longitude data were also transformed. The resulting coordinates preserve relative positions but not accurate real-world distances, preventing the identification of users' physical locations. The dataset additionally contains anonymized geographic block IDs mapped from the transformed coordinates.

\subsection{Multi-Task Sequence Construction}
The construction of the input sequence adopts a structure similar to that of IntSR~\cite{yan2025intsr}. Since our focus is on the recommendation recall problem, we omit the Q (query) tokens. Specifically, the input sequence comprises three types of tokens:

\begin{itemize}
    \item \textbf{S (Scenario tokens)}: These tokens encode the spatiotemporal context of user activities, allowing the model to learn context-aware user preferences.
    \item \textbf{I (Item tokens)}: These tokens form the chronological backbone of a user's interaction history. Each token corresponds to a POI the user has interacted with.
    \item \textbf{F (Feedback tokens)}: Paired with each item token, these tokens explicitly signal the user's interaction type, such as a ``click'' or an ``order''. These tokens provide crucial information about the intensity and type of user engagement, enriching the model's understanding of user intent.
\end{itemize}

\subsection{Multi-Task Label Attachment}

Multiple tasks share the same input sequence, while each task is supervised by a dedicated label sequence. To prevent data leakage in our chronologically ordered input, we strategically attach labels to specific token types based on the predictive goal of each task.

\begin{figure}[!t]
  \centering
  \includegraphics[width=0.6\columnwidth]{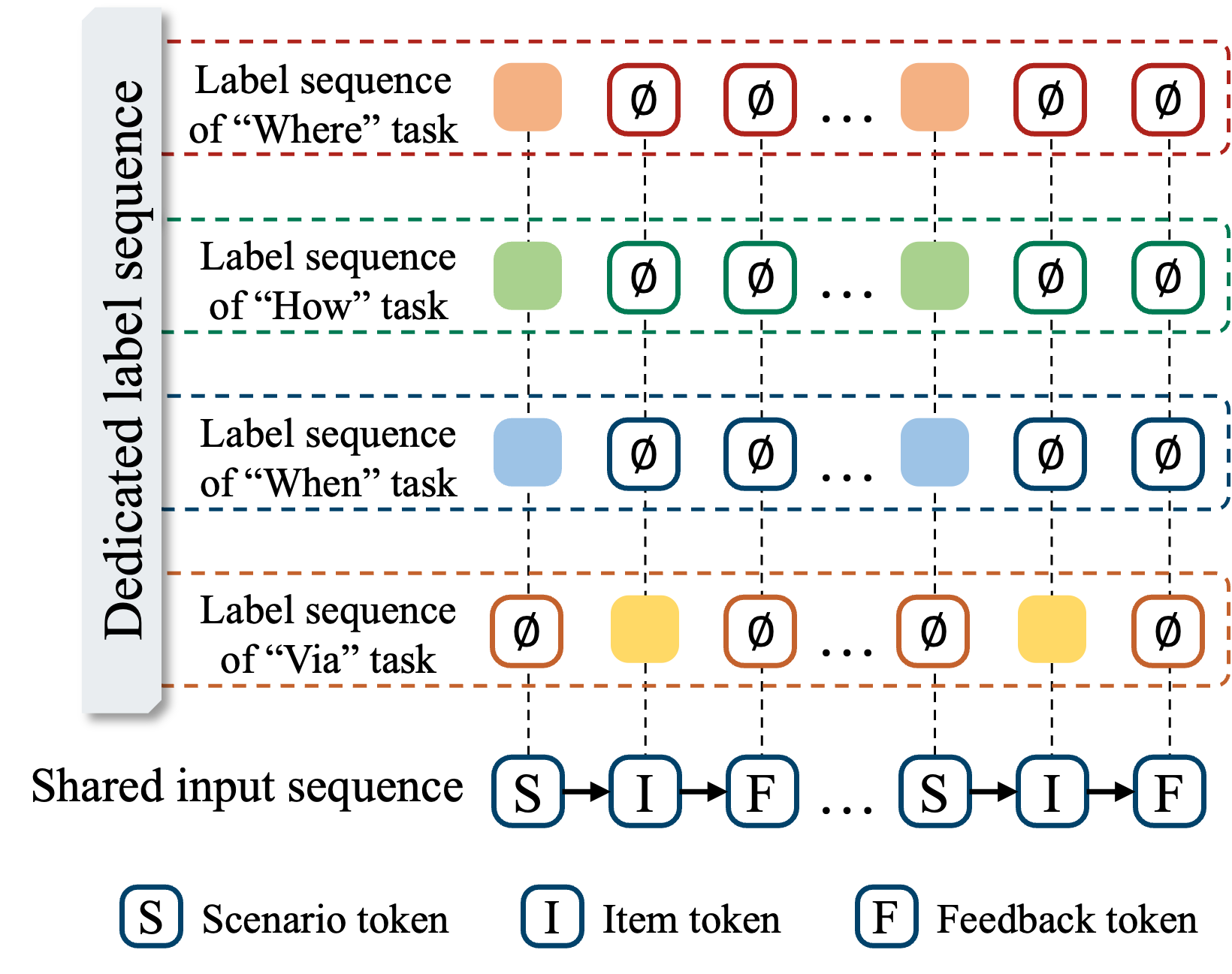}
  \caption{Tasks employ shared input sequence and dedicated label sequences.}
  \label{Figure_sequence_contruction}
\end{figure}

For the ``When,'' ``How,'' and ``Where'' tasks, labels are attached to the S tokens. This forces the model to make predictions using only the current context and past interactions. For the ``Via'' task, which focuses on predicting on-the-way intent, its labels are anchored to the I tokens, conditioning the prediction on a known destination. For any given task, tokens not relevant for its supervision are assigned a null label. This label attachment strategy is illustrated in Figure~\ref{Figure_sequence_contruction}. It works in conjunction with our decoder-only architecture and its inherent causal masking mechanism to strictly prevent information leakage.

\subsection{Dataset Splitting}
For the full IntTravel dataset, we adopted a temporal splitting strategy. For each user, all interactions excluding the last two were used to form the training set. The penultimate interaction was designated as the validation set, and the final interaction was used as the test set. Given the dataset's scale, the initial validation and test sets contained more than 160 million interactions each. We therefore sampled 0.5\% from each set, resulting in final validation and test sets of 815,000 interactions each.

\subsection{Prediction Targets and Candidate Construction}
\label{Section_candidate_construction}
The ground truth for the ``How'' task consists of six common travel modes, such as driving, public transport, and walking. For the ``When'' task, the ground truth spans a 48-hour range discretized into 48 hourly classes. No sampling is required for these two tasks because their candidate sets are small. For the ``Where'' and ``Via'' tasks, which involve millions of POIs, we employ a hybrid negative sampling strategy. For each interaction, 14 POIs are randomly sampled from the entire POI corpus, excluding the ground truth. We additionally sample up to 50 hard-negative POIs from the geographic block containing the ground-truth POI, excluding the ground truth itself; if fewer than 50 such POIs are available, all of them are included. Thus, each interaction has at most 64 negative samples.

\section{Preliminary}
Let $\mathcal{U}$ be the set of users and $\mathcal{I}$ be the set of POIs. $\mathcal{A}$ is the set of all user interactions. For each user $u \in \mathcal{U}$, their historical travel behavior is represented as a chronological sequence of interactions $\mathcal{A}_u = (i_1, i_2, \dots, i_N)$, where each interaction is associated with available contextual information, which may include a POI, a selected travel mode, and a timestamp. $N$ is the length of this sequence.

Let $\mathcal{K}$ be the set of all tasks. Our goal is to build a multi-task generative model that, given a user's historical interaction sequence, simultaneously predicts the recommendation result for each task $k \in \mathcal{K}$ for his/her next journey. All tasks share the same input sequence $\mathcal{A}_u$. The training objective is to find the optimal set of model parameters, $\Theta$, that maximizes the log-likelihood of observing the ground-truth labels for all tasks across the entire training dataset $\mathcal{D}$. This can be formally expressed as:
\begin{equation} \label{eq:objective}
    \hat{\Theta} = \argmax_{\Theta} \sum_{(\mathcal{A}, \mathbf{y}) \in \mathcal{D}} \log P([y_k|k \in \mathcal{K}] | \mathcal{A}; \Theta),
\end{equation}
where $y_k$ represents the ground-truth label of task $k$.

\section{IntTravel: Multi-Task Generative Framework}

\begin{figure}[t]
  \centering
  \includegraphics[width=\linewidth]{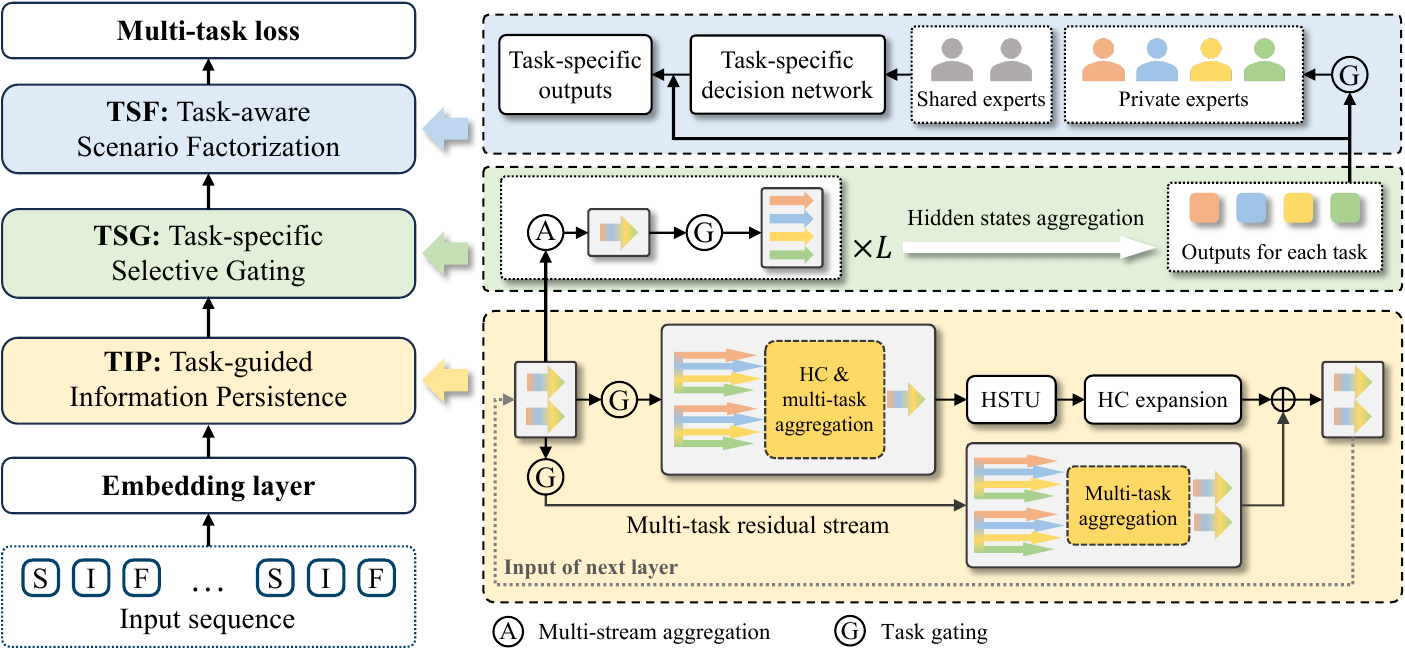}
  \caption{Multi-task framework of IntTravel. IntTravel stacks three core modules to handle multiple tasks. Task-Guided Information Persistence (TIP) ensures task-relevant information is retained. Task-Specific Selective Gating (TSG) filters useful information for each task. Task-Aware Scenario Factorization (TSF) generates task-aware parameters for the output.}
  \label{Figure_IntTravel_model}
\end{figure}

IntTravel proposes a bottom-up multi-task method to handle multiple tasks within a single generative model, as presented in Figure~\ref{Figure_IntTravel_model}. The approach comprises three modules: 

\begin{itemize}
    \item \textbf{Task-Guided Information Persistence (TIP)} ensures maximum propagation of task-relevant information in the decoder while keeping the computational complexity at $\mathcal{O}(N^2C)$.
    \item \textbf{Task-Specific Selective Gating (TSG)} enables each task to filter useful information from the decoder's output.
    \item \textbf{Task-Aware Scenario Factorization (TSF)} empowers each task to factorize its output based on specific scenarios.
\end{itemize}

\subsection{Task-Guided Information Persistence (TIP)}
\label{Section_TIP}

\subsubsection{HSTU}
Hierarchical Sequential Transduction Unit (HSTU) \cite{HSTU_zhai2024} is used as the sequence encoder in our IntTravel framework. It is a decoder-only architecture widely used in generative recommendation~\cite{yan2025intsr,Meituan_mtgr_han2025}. It stacks multiple uniform layers with residual connections. Each layer of HSTU is formulated in Eqs.~(\ref{Eq_TIP_HSTU_1})-(\ref{Eq_TIP_HSTU_2}), where $x_l, x_{l+1} \in \mathbb{R}^{N \times C}$ denote the input and output of the $l_{th}$ layer. $C$ is feature dimension. $U_l, V_l, Q_l$, and $K_l$ are gating weights, values, queries, and keys, respectively. Positional and temporal relative bias, $\text{rab}^{p} \in \mathbb{R}^{N \times N}$ and $\text{rab}^{t} \in \mathbb{R}^{N \times N}$, are introduced to refine the similarity scores. $f_1$ and $f_2$ are linear layers. Each layer of HSTU processes a single stream of input $x_l$. Causal mask is applied in HSTU to prevent information leakage. While the IntTravel framework can incorporate various sequence encoders such as HSTU and Multi-Head Attention (MHA), our primary contribution lies in the overall architecture rather than the specific encoder module.
\begin{equation}
    U_l, V_l, Q_l, K_l = \text{Split}(\text{SiLU}(f_1(x_l))),
    \label{Eq_TIP_HSTU_1}
\end{equation}
\begin{equation}
    x_{l+1} = f_2\left(\text{Norm}\left(\text{SiLU}\left(Q_l K_l^\top + \text{rab}^{p} + \text{rab}^{t}\right) V_l \right) \odot U_l \right).
    \label{Eq_TIP_HSTU_2}
\end{equation}

\subsubsection{TIP}
The TIP module is the multi-task version of Hyper Connection~(HC) embedded with HSTU, enabling both the advantages of residual connection expansion and the maximal persistence of task-aware information. HC has been validated as effective in expanding the width of residual streams without introducing much computation overhead~\cite{hyper_connection_zhu}. With the expansion rate denoted by $n$, the single residual stream $x_l$ is extended to a multi-stream representation $X_l \in \mathbb{R}^{N \times n \times C}$. The mechanism of HC can be defined as Eq.~(\ref{Eq_TIP_HC_common_formulation}). 
\begin{equation}
    X_{l+1} = H^{\mathrm{res}}_{l}\ X_{l}
    + \left(H^{\mathrm{post}}_{l}\right)^{\top} F \left(H^{\mathrm{pre}}_{l} X_{l}, W_l\right).
    \label{Eq_TIP_HC_common_formulation}
\end{equation}

$X_l$ and $X_{l+1}$ denote the input and output of the $l_{th}$ layer with HC. The initial hyper connection input is formed by replicating the network input for $n$ times, i.e., $X_0 = \left[x_0 \ x_0 \ ... \ x_0\right]^\top$. $H^{\mathrm{pre}}_{l} \in \mathbb{R}^{1 \times n}$ aggregates features from $n$ stream into one combined stream; $H^{\mathrm{post}}_{l} \in \mathbb{R}^{1 \times n}$ expands the output back to $n$ streams. $H^{\mathrm{res}}_{l} \in \mathbb{R}^{n \times n}$ serves as the mixing parameters of residual streams. $F$ represents any network layer, such as attention mechanisms or feed forward networks.

HC is static when $H^{\mathrm{pre}}_{l}, H^{\mathrm{post}}_{l}$, and $H^{\mathrm{res}}_{l}$ are initialized as learnable parameters. In dynamic HC, $H^{\mathrm{pre}}_{l}, H^{\mathrm{post}}_{l}$, and $H^{\mathrm{res}}_{l}$ are calculated as the summation of dynamic part and static part, as shown in Eq.~(\ref{Eq_TIP_dynamic_HC}). The static part is same as that of static HC (i.e., $b_l^{\text{pre}}, b_l^{\text{post}} \in \mathbb{R}^{1 \times n}, b_l^{\text{res}} \in \mathbb{R}^{n \times n}$), while the dynamic part depends on the input $X_l$. With the normalized input $\tilde{X}_l$, the dynamic part is derived by linear projections with parameters defined as $\theta_l^{\text{pre}}, \theta_l^{\text{post}} \in \mathbb{R}^{1 \times C}$ and $\theta_l^{\text{res}} \in \mathbb{R}^{n \times C}$. $\alpha_l^{\text{pre}}, \alpha_l^{\text{post}}, \alpha_l^{\text{res}} \in \mathbb{R}$ are learnable gating values.
\begin{equation}
    \left\{\begin{array}{l}           
        \tilde{X}_l = \mathrm{RMSNorm}(X_l) \\
        H_l^{\text{pre}} = \alpha_l^{\text{pre}} \cdot \tanh(\theta_l^{\text{pre}} \tilde{X}_l^\top) + b_l^{\text{pre}} \\
        H_l^{\text{post}} = \alpha_l^{\text{post}} \cdot \tanh(\theta_l^{\text{post}} \tilde{X}_l^\top) + b_l^{\text{post}} \\
        H_l^{\text{res}} = \alpha_l^{\text{res}} \cdot \tanh(\theta_l^{\text{res}} \tilde{X}_l^\top) + b_l^{\text{res}}
    \end{array} \right ..
    \label{Eq_TIP_dynamic_HC}
\end{equation}

HC is not designed for multi-task problems, and thus, is not capable of handling multiple tasks. We extend the standard HC to a multi-task version by incorporating a task-gating mechanism. As formulated in Eq.~(\ref{Eq_TIP_H_pre_res_taskgate}), the task-specific parameters, $J^{\mathrm{pre}}_{l}  \in \mathbb{R}^{K \times 1 \times n}$ and $J^{\mathrm{res}}_{l} \in \mathbb{R}^{K \times n \times n}$, are generated for each task $k$, where $K=|\mathcal{K}|$ is the number of tasks. This is achieved by performing an element-wise Hadamard product ($\odot$) between the primary, task-agnostic parameters~($H^{\mathrm{pre}}_{l}$ and $H^{\mathrm{res}}_{l}$) and a corresponding task- and stream-specific scalar gate, $\gamma_{l,k}$. This gate acts as a dynamic filter, selectively scaling the shared primary parameters to create a customized set of parameters tailored to the specific needs of task $k$.
\begin{equation}
    J^{\mathrm{pre}}_{l} = H^{\mathrm{pre}}_{l} \odot \gamma_{l,k}, J^{\mathrm{res}}_{l} = H^{\mathrm{res}}_{l} \odot \gamma_{l,k}.
    \label{Eq_TIP_H_pre_res_taskgate}
\end{equation}

Therefore, the multi-task version of HC with $F$ instantiated as HSTU can be formulated as:
\begin{equation}
    X_{l+1} = \text{Agg}_{k} \left(J^{\mathrm{res}}_{l} X_{l} \right)
    + \left(H^{\mathrm{post}}_{l}\right)^{\top} \mathrm{HSTU} \left(\text{Agg}_{k} \left(J^{\mathrm{pre}}_{l} X_{l} \right), W_l\right),
    \label{Eq_TIP_HC_multi_task}
\end{equation}

where $\text{Agg}_{k}(\cdot)$ denotes an aggregation function to combine the information learned across all tasks. For example, a simple aggregation function could be to sum or average pooling over the task dimension $k$. In IntTravel, average pooling is used here. In this way, the final results contain the collective information from all individual tasks.

\subsection{Task-Specific Selective Gating (TSG)}
The purpose of TSG is to select task-specific information and to form the input of following scenario factorization network. The process involves three key steps. First, to create a unified representation for each decoder layer's output, the multi-stream output $X_l \in \mathbb{R}^{N \times n \times C}$ from the preceding TIP layer is aggregated along stream dimension to obtain $z_{l} \in \mathbb{R}^{N \times C}$, see Eq.~(\ref{Eq_TSK_stream_aggregation}). In IntTravel, $\text{Agg}_{n}(\cdot)$ is defined as sum pooling followed by layer normalization.
\begin{equation}
    z_{l} = \text{Agg}_{n}(X_l).
    \label{Eq_TSK_stream_aggregation}
\end{equation}

Second, to enable task-specific filtering, the module must determine the relevance of each layer's information for each task. To this end, a series of layer-specific gating scalars, $s_{l,k}$, are generated for each layer $l$ and each task $k$ from its unique task embedding $\text{Emb}_k$ via a layer-specific MLP, as shown in Eq.~(\ref{Eq_TSG_layer_specific_task_gate}). $s_{l,k}$ learns to quantify the importance of layer $l$'s output for task $k$. With these importance scores, the module performs the selective gating. As formulated in Eq.~(\ref{Eq_TSG_task_specific_selective_gate}), $z_l$ is scaled by $s_{l,k}$ through element-wise product. This operation effectively filters the layer's information by amplifying features relevant to task $k$ and suppressing irrelevant ones, resulting in a task-aware representation $Z_{l} \in \mathbb{R}^{N \times K \times C}$.
\begin{equation}
    s_{l,k} = \text{MLP}(\text{Emb}_k),
    \label{Eq_TSG_layer_specific_task_gate}
\end{equation}
\begin{equation}
    Z_{l,k} = s_{l,k} \odot z_l.
    \label{Eq_TSG_task_specific_selective_gate}
\end{equation}

Finally, to construct a single, comprehensive output for each task, the filtered, layer-specific representations $Z_{l,k}$ are aggregated across all decoder layers $l$. This step combines the selected information from the entire decoder depth into the final task-specific output of the TSG module, $T_k$, as shown in Eq.~(\ref{Eq_TSG_layer_aggregation}). When multiple decoder layers are used, this aggregation concatenates the filtered representation from the last decoder layer with the average of the filtered representations from all preceding layers. For a single-layer model, the only filtered representation is used directly.
\begin{equation}
    T_k =
    \begin{cases}
        Z_{1,k}, & L=1, \\
        \text{concat}\left(Z_{L,k}, \dfrac{1}{L-1}\displaystyle\sum_{l=1}^{L-1} Z_{l,k}\right), & L>1,
    \end{cases}
    \label{Eq_TSG_layer_aggregation}
\end{equation}
where $L$ denotes the number of decoder layers, $Z_{l,k}$ denotes the filtered representation of task $k$ at layer $l$, and $T_k \in \mathbb{R}^{N \times C_T}$, with $C_T$ denoting the output feature dimension of TSG.

\subsection{Task-Aware Scenario Factorization (TSF)}

The Task-Aware Scenario Factorization (TSF) module is designed to handle different tasks and user preferences. This is achieved by dynamically creating customized MLP layers that adapt to each specific situation. Inspired by the concept of scenario-based parameter generation of DSFNet~\cite{yu2025dsfnet}, we enhance this idea by introducing a clear distinction between shared and task-private experts for better multi-task learning. Specifically, we define two types of experts: 
\begin{itemize}
    \item A set of \textbf{shared experts}, $\mathcal{E}^s$, which contain general knowledge that is useful for all tasks.
    \item A set of \textbf{private experts}, $\mathcal{E}_k^p$, which hold specialized knowledge of each individual task $k$.
\end{itemize}

For each task $k$, we construct a task-specific context representation $R_k$ by concatenating the task embedding $\text{Emb}_k$ with user profile features $p$~(Eq.~(\ref{Eq_TSF_concat_profile})). $R_k$ is passed through an MLP with sigmoid function $\sigma$ and multiplied by 2 to obtain $\beta_{e}$~(Eq.~(\ref{Eq_TSF_expert_weights})). This allows the weight to be greater than 1, enabling the model to amplify the influence of experts that are particularly important for the current context. To decide how much each expert should contribute for a given situation, we calculate the weight vector $\beta \in \mathbb{R}^{N \times E}$, where $E=|\mathcal{E}^s \cup \mathcal{E}^p_k|$ is the number of experts for task $k$.
\begin{equation}
    R_k=\text{concat}(\text{Emb}_k, p),
    \label{Eq_TSF_concat_profile}
\end{equation}
\begin{equation}
    \beta = 2 * \sigma \left( \text{MLP} (R_k) \right).
    \label{Eq_TSF_expert_weights}
\end{equation}

The final parameters of task $k$, $W_k \in \mathbb{R}^{C_T \times C'}$ and $b_k \in \mathbb{R}^{C'}$, are calculated as the weighted sum of the base parameters ($\tilde{W}_{e} \in \mathbb{R}^{C_T \times C'}$, $\tilde{b}_{e} \in \mathbb{R}^{C'}$) from a selected group of experts~(Eq.~(\ref{Eq_TSF_task_specific_mlp_parameters})), where $C'$ is the output feature dimension of the TSF layer.

For task $k$, we only sum over the shared experts ($\mathcal{E}^s$) and that specific task's private experts ($\mathcal{E}_k^p$). This allows each task to effectively combine general, shared knowledge with its own unique, specialized knowledge. In this way, the TSF module constructs an MLP layer whose parameters are customized to both the specific task $k$ and the current user profile $p$.

In parallel, we refine the task-specific input features $T_k$ using a filtering mechanism. As shown in Eq.~(\ref{Eq_TSF_feature_filtering}), a dynamic gate is generated from both $T_k$ and the context representation $R_k$. This gate is then applied element-wise to $T_k$ to produce the filtered features $\tilde{T}_k$. Multiple layers of TSF can be stacked to enhance modeling capacity. Taking one layer of TSF as an example, the output of this TSF layer for task $k$, $O_k \in \mathbb{R}^{N \times C'}$, is obtained by applying this newly constructed MLP layer (with parameters $W_k$ and $b_k$) to its corresponding input representation, $\tilde{T}_k$, as shown in Eq.~(\ref{Eq_TSF_task_specific_output}).
\begin{equation}
    W_k = \sum_{e \in \mathcal{E}^s \cup \mathcal{E}^p_k} \beta_{e}\tilde{W}_{e},
    b_k = \sum_{e \in \mathcal{E}^s \cup \mathcal{E}^p_k} \beta_{e}\tilde{b}_{e},
    \label{Eq_TSF_task_specific_mlp_parameters}
\end{equation}
\begin{equation}
    \tilde{T}_k = T_k \odot \sigma (\text{MLP}(\text{concat}(T_k, R_k))),
    \label{Eq_TSF_feature_filtering}
\end{equation}
\begin{equation}
    O_k = \tilde{T}_k W_k + b_k.
    \label{Eq_TSF_task_specific_output}
\end{equation}

\subsection{Loss Function}
For each task $k$, we compute an InfoNCE loss that maximizes the probability of the ground-truth candidate, $\omega^+$, over the task-specific candidate set $\Omega_{u,a,k}$. For the ``When'' and ``How'' tasks, $\Omega_{u,a,k}$ contains all departure-time classes and travel-mode classes, respectively. For the ``Where'' and ``Via'' tasks, it contains the ground-truth POI together with sampled negative POIs. Let $\mathcal{A}_{u,k} \subseteq \mathcal{A}_u$ denote the interactions with a non-null label for task $k$, and let $|\mathcal{A}_k|=\sum_{u \in \mathcal{U}}|\mathcal{A}_{u,k}|$. Interactions with null labels are excluded from the loss computation. The logit, $\hat{y}^{\omega}_k$, is calculated as the dot product of $O_k$ and the embedding of candidate $\omega$. The final multi-task loss $L$ is then the sum of all individual task losses, as shown in Eq.~(\ref{Eq_loss}).
\begin{equation}
    L = \sum_{k \in \mathcal{K}} -\frac{1}{|\mathcal{A}_k|} \sum_{u \in \mathcal{U}} \sum_{a \in \mathcal{A}_{u,k}} \text{log} \frac{\text{exp}(\hat{y}^{\omega^+}_k)}{\sum_{\omega \in \Omega_{u,a,k}}\text{exp}({\hat{y}^{\omega}_k})}.
    \label{Eq_loss}
\end{equation}

\section{Offline Evaluation}

\subsection{Baselines}

We compare IntTravel against the following multi-task learning baselines: \textbf{STEMNet}~\cite{stem_su2024}, \textbf{MuSeNet}~\cite{musenet_xu2023}, \textbf{APG}~\cite{apg_yan2022}, \textbf{HiNet}~\cite{hinet_zhou2023}, \textbf{M2M}~\cite{M2M_zhang2022leaving}, \textbf{PLE}~\cite{ple_tang2020progressive}, \textbf{STAR}~\cite{star_sheng2021one}, and \textbf{HoME}~\cite{home_wang2025}.

\subsection{Implementation Details}

We set the embedding size to 96, the maximum sequence length to 120, and the batch size to 64. The model was trained using a learning rate of $1 \times 10^{-3}$. Unless otherwise specified (as in the scaling law experiments), the number of layers in TIP module was set to 3. All baseline models adopted these same hyperparameter settings. For TSF module, we set 2 shared experts and 1 private expert for each task. When implementing baseline models, we set the decoder to the original HSTU with the same number of layers. The TSG module was removed, and the TSF module was replaced by the respective baseline models, which served as the multi-task heads. All models are trained using Adam optimizer~\cite{adam2014method} on 8 NVIDIA H20 GPUs with 96 GB memory. All experimental results are averaged over 5 random seeds.

\subsection{Evaluation Metrics}

Evaluation metrics of the four tasks are detailed in Appendix~\ref{Appendix_evaluation_metrics}, including positive indicators such as Accuracy (Acc) and Hit Rate (HR), and negative ones like Mean Absolute Error (MAE), Bad Case Rate (BCR), and Category Inconsistency Rate (CIR). The HR metrics of ``Where'' and ``Via'' tasks are calculated against the negative samples described in Sec.~\ref{Section_candidate_construction}.\footnote{The offline performance gains in the "where" and "via" tasks were highly consistent with the improvements observed in our large-scale online A/B tests.} We report top-1 and top-5 HR.

\subsection{Overall Comparisons}

As detailed in Table~\ref{Table_result_comparison_our_dataset}, IntTravel achieves the best performance on every metric across all four tasks, with statistically significant improvements over the strongest baselines on the ``How'' task (paired $t$-test, $p < 0.05$, averaged over 5 random seeds). These results highlight the effectiveness of our architecture in capturing the relationships among travel-related predictions. Components like TIP and TSF allow the model to maintain crucial context and generate customized parameters for each specific travel scenario.

\begin{table}[htbp]
  \centering
  \caption{Comparison with baseline models across all four travel tasks on IntTravel dataset. For metrics, ↑ indicates higher is better, and ↓ indicates lower is better. The best results are highlighted with a \colorbox{dreamblue!15}{blue background}, and the second-best results are in \textcolor{dreamblue}{blue}. $^*$ indicates a statistically significant improvement of IntTravel over the strongest baseline (paired $t$-test, $p < 0.05$, averaged over 5 random seeds).}
  \label{Table_result_comparison_our_dataset}
  \setlength{\tabcolsep}{3pt}
\resizebox{\textwidth}{!}{%
\begin{tabular}{l cc cc ccc ccc}
    \toprule
    \multirow{2}{*}{\textbf{Model}} & \multicolumn{2}{c}{\textbf{``When'' task}} & \multicolumn{2}{c}{\textbf{``How'' task}} & \multicolumn{3}{c}{\textbf{``Where'' task}} & \multicolumn{3}{c}{\textbf{``Via'' task}} \\
    \cmidrule(lr){2-3} \cmidrule(lr){4-5} \cmidrule(lr){6-8} \cmidrule(lr){9-11}
    & Acc ↑ & MAE ↓ & Acc ↑ & BCR ↓ & HR@1 ↑ & HR@5 ↑ & CIR ↓ & HR@1 ↑ & HR@5 ↑ & CIR ↓ \\
    \midrule
    STEM-Net  & 0.8328 & 8.0249 & 0.6697 & 0.0730 & 0.6582 & 0.8582 & 0.2667 & 0.6633 & 0.8606 & 0.2667 \\
    MuSeNet  & 0.8326 & 8.0344 & 0.6658 & 0.0736 & 0.6732 & \second{0.8643} & 0.2548 & 0.6823 & 0.8669 & \second{0.2522} \\
    APG      & 0.8321 & 8.0609 & 0.6697 & 0.0733 & 0.6429 & 0.8524 & 0.2779 & 0.6459 & 0.8545 & 0.2792 \\
    HiNet    & 0.8326 & 8.0349 & 0.6666 & 0.0746 & 0.6254 & 0.8461 & 0.2919 & 0.6266 & 0.8467 & 0.2949 \\
    M2M      & 0.8328 & 8.0239 & 0.6675 & 0.0740 & 0.6583 & 0.8587 & 0.2653 & 0.6676 & 0.8625 & 0.2624 \\
    PLE      & 0.8326 & 8.0339 & \second{0.6706} & \second{0.0727} & 0.6579 & 0.8577 & 0.2666 & 0.6633 & 0.8599 & 0.2667 \\
    STAR     & \second{0.8329} & \second{8.0229} & 0.6696 & 0.0733 & \second{0.6749} & 0.8642 & \second{0.2545} & \second{0.6832} & \second{0.8674} & 0.2525 \\
    HoME     & 0.8326 & 8.0344 & 0.6694 & 0.0734 & 0.6585 & 0.8581 & 0.2663 & 0.6644 & 0.8603 & 0.2656 \\
    \midrule
    IntTravel & \best{0.8332} & \best{8.0077} & \best{0.6734}$^*$ & \best{0.0717}$^*$ & \best{0.6840} & \best{0.8693}$^*$ & \best{0.2460} & \best{0.6924} & \best{0.8722} & \best{0.2449} \\
    \bottomrule
  \end{tabular}
}
\end{table}

\subsection{Ablation Studies}

To validate the effectiveness of each component in IntTravel, we conduct extensive ablation studies. We compare the \textbf{full version} of IntTravel against several variants with specific modules removed or replaced:

\begin{itemize}
    \item \textbf{Ablation of TIP}: We evaluate the contribution of TIP module by ablating its core components in following four variants: (1)~\textit{w/o} $J^{\text{pre}}$: $J^{\text{pre}}$ is replaced with $H_{\text{pre}}$. (2)~\textit{w/o} $J^{\text{res}}$: $J^{\text{res}}$ is replaced with $H_{\text{res}}$. (3)~\textit{w/o} $J^{\text{pre}}$ and $J^{\text{res}}$: $J^{\text{pre}}$ and $J^{\text{res}}$ are replaced with $H_{\text{pre}}$ and $H_{\text{res}}$, respectively. (4)~\textit{w/o} TIP: the TIP module is replaced with the original version of HSTU without multi-task HC.

    \item \textbf{Ablation of TSG}: We investigate the two main mechanisms within TSG module by the following ablation settings: (1)~\textit{w/o hidden states}: Removes the aggregation of intermediate hidden states from the HSTU layers, relying exclusively on the final layer's output. (2)~\textit{w/o task gating}: Disables the task-specific gating mechanism, forcing all tasks to share the same fused representation without differentiation. The input of TSF is the average pooling of all HSTU layers' outputs.

    \item \textbf{Ablation of TSF}: \textit{w/o TSF}: the entire TSF module is replaced with task-specific MLP layers.
\end{itemize}

\begin{table}[htbp]
    \centering
    \caption{Results of ablation experiments. For metrics, ↑ indicates higher is better, and ↓ indicates lower is better. The best results are highlighted with a \colorbox{dreamblue!15}{blue background}, and the second-best results are in \textcolor{dreamblue}{blue}.}
    \label{Table_ablation_study}
    \begingroup
    \setlength{\tabcolsep}{3pt}
\resizebox{\textwidth}{!}{%
\begin{tabular}{ll cc cc ccc ccc}
        \toprule
        \multicolumn{2}{l}{\multirow{2}{*}{\textbf{Model}}} & \multicolumn{2}{c}{\textbf{``When'' task}} & \multicolumn{2}{c}{\textbf{``How'' task}} & \multicolumn{3}{c}{\textbf{``Where'' task}} & \multicolumn{3}{c}{\textbf{``Via'' task}} \\
        \cmidrule(lr){3-4} \cmidrule(lr){5-6} \cmidrule(lr){7-9} \cmidrule(lr){10-12}
        \multicolumn{2}{l}{} & Acc ↑ & MAE ↓ & Acc ↑ & BCR ↓ & HR@1 ↑ & HR@5 ↑ & CIR ↓ & HR@1 ↑ & HR@5 ↑ & CIR ↓ \\
        \midrule
        \multirow{4}{*}{TIP} & \textit{w/o} $J^{\text{pre}}$ & 0.8328 & 8.0256 & 0.6706 & 0.0726 & 0.6806 & 0.8681 & 0.2490 & 0.6890 & 0.8709 & 0.2476 \\
        & \textit{w/o} $J^{\text{res}}$ & 0.8327 & 8.0284 & \second{0.6724} & \second{0.0720} & 0.6818 & \second{0.8685} & 0.2480 & 0.6901 & \second{0.8715} & 0.2470 \\
        & \textit{w/o} $J^{\text{pre}}$ and $J^{\text{res}}$ & 0.8328 & 8.0265 & 0.6719 & 0.0720 & \second{0.6832} & 0.8684 & \second{0.2470} & \second{0.6911} & 0.8710 & \second{0.2461} \\
        & \textit{w/o} TIP & 0.8328 & 8.0256 & 0.6688 & 0.0737 & 0.6741 & 0.8653 & 0.2540 & 0.6828 & 0.8684 & 0.2519 \\
        \midrule
        \multirow{2}{*}{TSG} & \textit{w/o} hidden states & 0.8327 & 8.0299 & 0.6704 & 0.0730 & 0.6751 & 0.8658 & 0.2537 & 0.6834 & 0.8685 & 0.2522 \\
        & \textit{w/o} task gating & 0.8326 & 8.0337 & 0.6675 & 0.0738 & 0.6769 & 0.8669 & 0.2522 & 0.6865 & 0.8704 & 0.2497 \\
        \midrule
        TSF & \textit{w/o} TSF & \second{0.8329} & \second{8.0217} & 0.6692 & 0.0728 & 0.6803 & 0.8658 & 0.2494 & 0.6886 & 0.8682 & 0.2483 \\
        \midrule
        \multicolumn{2}{l}{Full version} & \best{0.8332} & \best{8.0077} & \best{0.6734} & \best{0.0717} & \best{0.6840} & \best{0.8693} & \best{0.2460} & \best{0.6924} & \best{0.8722} & \best{0.2449} \\
        \bottomrule
    \end{tabular}
}
    \endgroup
\end{table}

The results presented in Table~\ref{Table_ablation_study} confirm the critical contribution of each proposed module. Removing TIP yields the largest degradation in ``Where'' HR@1, ``Where'' HR@5, ``Where'' CIR, and ``Via'' HR@1, indicating that propagating and preserving task-relevant information across layers is crucial. Removing TSF also reduces performance on the POI-retrieval tasks, underscoring the importance of dynamically generating customized parameters for each specific task-user scenario. Furthermore, ablating the gating mechanism within TSG produces the lowest ``How'' accuracy and the highest ``How'' BCR, highlighting the necessity of its selective filtering capability. The full version model effectively combines these modules to achieve the best overall results.

\subsection{Scaling Laws}

\begin{figure}[htbp]
    \centering
    \includegraphics[width=0.6\columnwidth]{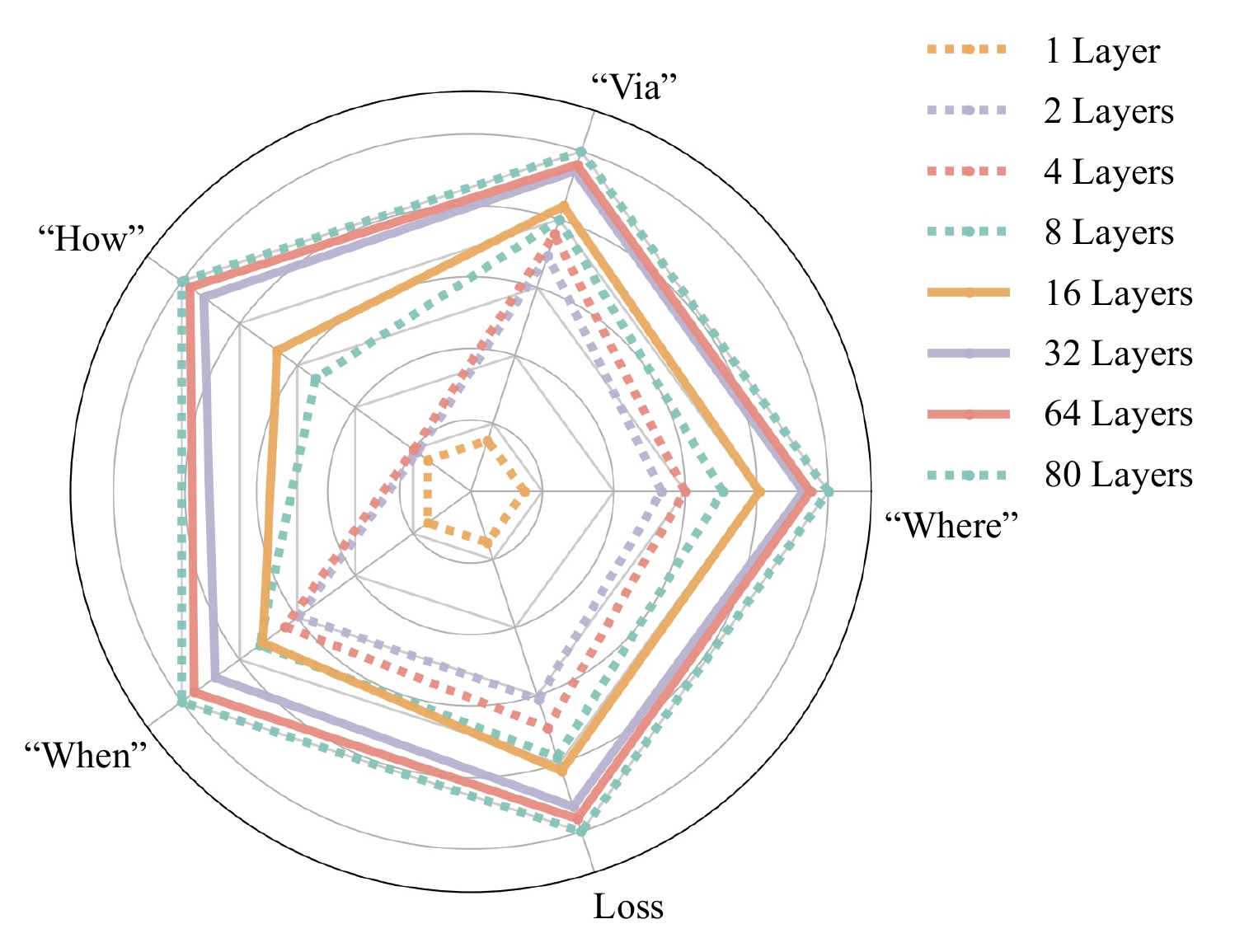}
    \caption{Scaling effects with varying layers. All plotted points are computed after min–max normalization. Task performance increases outward while loss decreases outward.}
    \label{Figure_scaling}
\end{figure}

To investigate the scaling capability of our model, we trained several models, with depths ranging from 1 layer to 80 layers. The performance of these variants was evaluated with positive metrics of all four tasks~(Acc for ``When'' and ``How'' tasks, and HR@1 for ``Where'' and ``Via'' tasks), as well as the total loss. The results are visualized in Figure~\ref{Figure_scaling}, revealing a strong and consistent scaling trend. As the model depth increases from 1 to 80 layers, performance on all metrics steadily improves. There is no sign of performance decrease even at very deep configurations like 80 layers, which is often a challenge for complex models. This demonstrates the excellent scaling capability of proposed architecture.

\section{Online Deployment}
\subsection{System Architecture}

\begin{figure}[htbp]
  \begin{center}
    \centerline{\includegraphics[width=0.7\columnwidth]{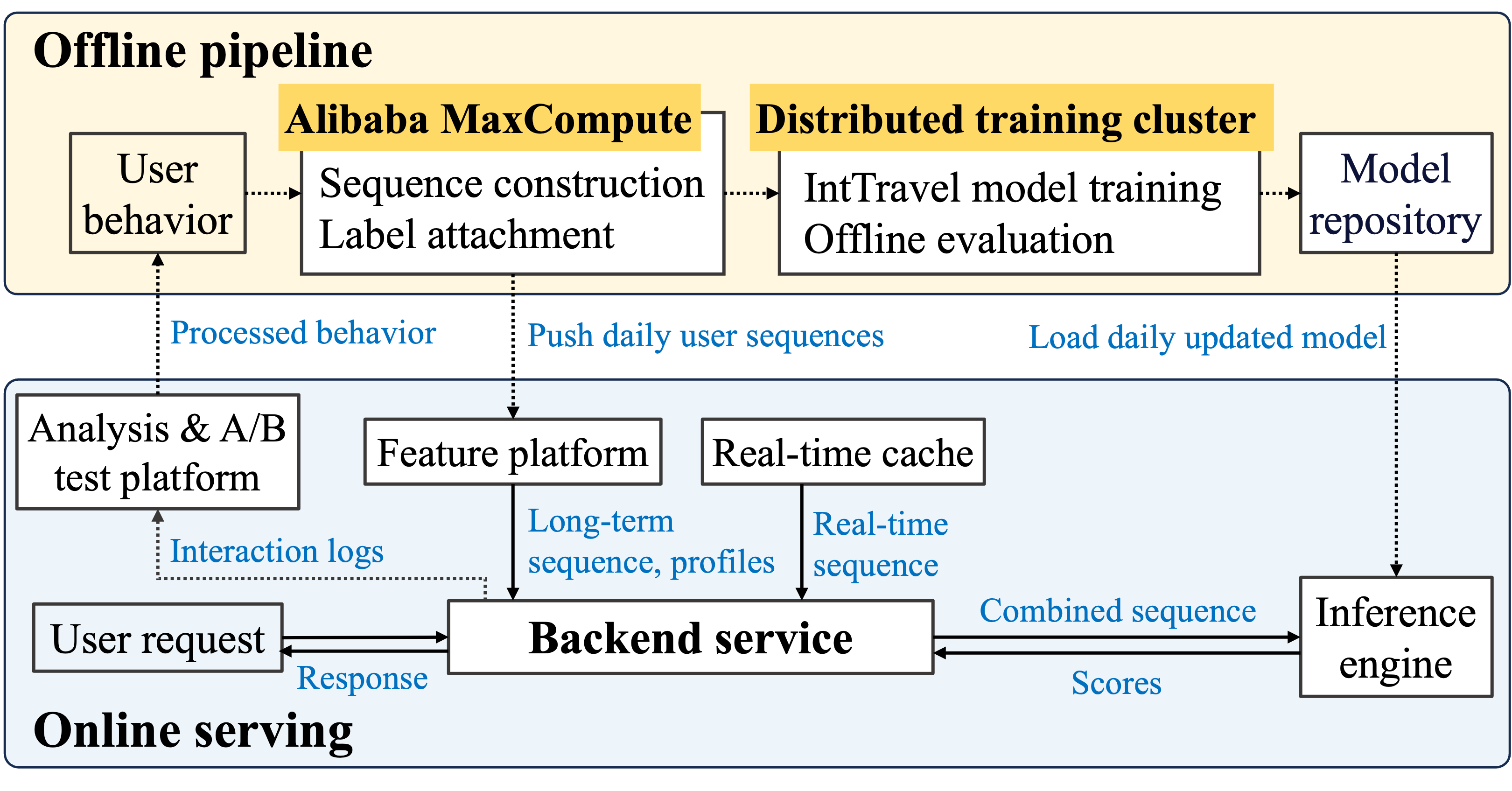}}
    \caption{
      The end-to-end system architecture of IntTravel.
    }
    \label{Offline_pipeline_and_online_serving}
  \end{center}
\end{figure}

Our system architecture (Figure~\ref{Offline_pipeline_and_online_serving}) is designed for industrial-scale deployment, balancing daily model updates with real-time serving latency. It comprises an offline pipeline and an online serving flow.

\textbf{Offline pipeline}: A daily pipeline processes raw user logs (clicks, searches, orders, route plans, and navigation) into chronological interaction sequences. These sequences, serving as our primary features, are precomputed and pushed to an online key-value feature platform to minimize online latency. The updated IntTravel model is trained on these sequences and deployed via a model repository, ensuring seamless version updates for the online engines.

\textbf{Online serving flow}: Upon receiving a user request, the backend service fetches the precomputed historical sequence from the feature platform and merges it with the user's real-time interactions from an in-memory cache. This complete sequence is then sent to the inference engine. For the ``Where'' and ``Via'' tasks, the model's outputs are used to retrieve a set of POIs via a retrieval service based on Approximate Nearest Neighbor~(ANN) algorithm. The final recommendation results are returned to the user, and their feedback is logged to close the training loop. This architecture ensures both personalization and responsiveness at a massive scale.

\subsection{Online A/B Test }
We conducted an online A/B test of the IntTravel multi-task model within the ``Nearby POI Recommendation'' feature of the Amap application. The model was compared against the existing online baseline, IntSR~(non-multi-task version of IntTravel), with each experimental group being allocated 5\% of the user traffic. The results of the two-week A/B test showed that IntTravel led to a \textbf{1.09\% relative increase} in CTR~(details in Table~\ref{table_user_segment_results}), demonstrating its significant value in a real-world production environment.

\begin{table}[htbp]
  \centering
  \caption{CTR uplift (relative) for different user segments. All results are statistically significant ($p < 0.0001$).}
  \label{table_user_segment_results}
  \begin{tabular}{lcc}
    \toprule
    \textbf{User segment (interactions/week)} & \textbf{Share} & \textbf{CTR lift} \\
    \midrule
    High-frequency (> 5) & 53\% & +1.088\% \\
    Medium-frequency (2--5) & 25\% & +1.106\% \\
    Low-frequency ($\le$ 1) & 22\% & +1.040\% \\
    \bottomrule
  \end{tabular}
\end{table}

\subsection{Online Efficiency}
Regarding online efficiency, our model handles a sequence length of 300 (covering 100 most recent behavioral interactions) under a peak concurrency of 30k QPS (query per second). Deployed on a cluster of 150 NVIDIA H20 GPUs (96 GB memory), the average response time (RT) is 40 ms and the 99\% RT is 100 ms. These metrics demonstrate that IntTravel is highly optimized and fully capable of supporting large-scale, high-concurrency industrial applications.

\section{Related Works}
\subsection{Travel Datasets}
Travel planning is a complex real-world challenge involving POI discovery, transportation and route optimization, and the creation of short- and long-distance travel plans. Recent research uses AI agents to manage constraints, decompose objectives, and optimize plans through iterative refinement and tool integration. Corresponding datasets and benchmarks~\citep{xie2024travelplanner,wang2025triptailor,hao2025large} have been developed to evaluate agent performance. In many real-world travel applications, however, recommendation requires predicting a user's next journey across four fundamental dimensions: departure time (When), transportation mode (How), destination (Where), and on-the-way intent (Via). Existing next-POI datasets~\citep{yang2014modeling,cho2011friendship,cheng2011exploring} primarily support destination prediction and do not capture the modal and situational dimensions of a journey. Their limited scale also constrains reliable evaluation of modern generative recommendation models.

\subsection{Multi-Task Learning}
Multi-task learning models have evolved significantly, moving from simple shared structures to more adaptive architectures. Early models, such as Shared-Bottom~\cite{sharedbottom_caruana1997multitask}, used a hard parameter sharing approach with common bottom layers. However, this often led to negative transfer when tasks were not closely related. To address this, soft parameter sharing models emerged. MMoE~\cite{mmoe_ma2018modeling} introduced task-specific gates to flexibly combine shared experts. PLE~\cite{ple_tang2020progressive} further improved this by explicitly separating shared and task-specific experts, effectively mitigating the ``seesaw phenomenon.'' As systems grew to serve multiple scenarios, the focus expanded. Models like STAR~\cite{star_sheng2021one}, M2M~\cite{M2M_zhang2022leaving}, and HiNet~\cite{hinet_zhou2023} developed specialized architectures to explicitly manage inter-scenario relationships. More recently, the granularity of sharing has advanced to the embedding level. STEM-Net~\cite{stem_su2024} pioneers learning both shared and task-specific embeddings to capture diverse user preferences.

\subsection{Generative Recommendation}

Generative Recommendation (GR) transforms traditional retrieval into sequence generation~\citep{geng2022recommendation,zheng2024adapting,liu2025generative,xu2025climber}. In contrast to traditional deep-learning-based models, generative recommenders possess the potential to alleviate cascaded error propagation and facilitate optimization across broader user behavioral contexts. TIGER~\citep{Tiger_rajput2023} first predicted hierarchical SIDs via a sequence-to-sequence model, while HSTU~\citep{HSTU_zhai2024} unified retrieval and ranking by interleaving item and action tokens. IntSR~\citep{yan2025intsr} further integrated search and recommendation through a query-driven block that isolates query placeholders from user history. OneRec~\citep{OneRec_deng2025} replaced multi-stage pipelines with an end-to-end generative architecture. However, research addressing multi-task learning within the generative paradigm remains scant.

\section{Conclusions}

In this paper, we presented IntTravel, a scalable generative multi-task system designed to solve a core industrial challenge at Amap: unifying fragmented recommendation services into a single framework. Our work demonstrates successful deployment of an end-to-end generative paradigm that holistically models a user's complete travel intent including ``when, where, how, and via''. We designed a novel architecture (TIP, TSG, TSF) that resolves the incompatibility between traditional multi-task methods and modern generative models. The business value of this approach was confirmed through online A/B testing, where IntTravel delivered a significant 1.09\% relative increase in CTR over a strong baseline. As part of our contribution, we have also released the large-scale IntTravel dataset to bridge the gap between academic research and industrial practice. Its scale and joint coverage of four travel tasks provide a resource for studying the interdependencies among departure time, travel mode, destination, and via-POI intent.

\newpage
\bibliographystyle{iclr2025_conference}
\bibliography{iclr2025_conference}

\newpage
\appendix
\section{IntTravel Dataset: Fields, Statistics, and Examples}
\label{Appendix_data_statistics_and_examples}

\subsection{Information of POIs}

The IntTravel dataset contains 7,291,872 POIs distributed across 10 major cities in China. Each POI is described by the following fields (Table~\ref{Table_POI_data_examples}):

\begin{itemize}
    \item \textbf{POI ID}: A unique identifier assigned to each POI. These IDs are anonymized by randomly shuffling the original identifiers and re-indexing them from 0.
    \item \textbf{Normalized score (Nscore)}: A score ranging from 0 to 1 that reflects the overall popularity of a POI. This score is derived from multiple factors, including interaction frequency. The specific calculation formula is not disclosed, and users can treat this score as a popularity-like feature.
    \item \textbf{Geographic ID (GID)}: An identifier for the geographic block where the POI is located. These block IDs have been anonymized and re-indexed. Adjacent IDs do not imply geographic proximity, but POIs sharing the same GID can be considered geographically close.
    \item \textbf{Category ID (CID)}: An integer identifier representing the POI category. For data protection, the original category labels have been converted into numerical IDs starting from 0.
    \item \textbf{Administrative Region ID (ARID)}: An integer identifier for the POI's administrative region. The original region identifiers have been transformed into numerical IDs to preserve privacy.
    \item \textbf{Coordinates}: The provided $(x,y)$ coordinates are anonymized spatial locations projected onto a two-dimensional Cartesian plane. They are not standard longitude and latitude values but preserve the relative positions of the original data.
\end{itemize}

\begin{table}[htbp]
\centering
\caption{Examples of POI information.}
\label{Table_POI_data_examples}
\small
\begin{tabular}{cccccc}
    \toprule
    \textbf{POI ID} & \textbf{Nscore} & \textbf{GID} & \textbf{CID} & \textbf{ARID} & \textbf{Coordinates} \\
    \midrule
    0 & 0.436961 & 25556599 & 8 & 353 & 21541.38, -16385.12 \\
    1 & 0.389202 & 5901606 & 1 & 2089 & 21543.91, -16389.45 \\
    2 & 0.190659 & 17952948 & 17 & 3192 & 21520.76, -16412.67 \\
    \bottomrule
\end{tabular}
\end{table}

\begin{figure*}[htbp]
    \centering
    \includegraphics[width=0.95\linewidth]{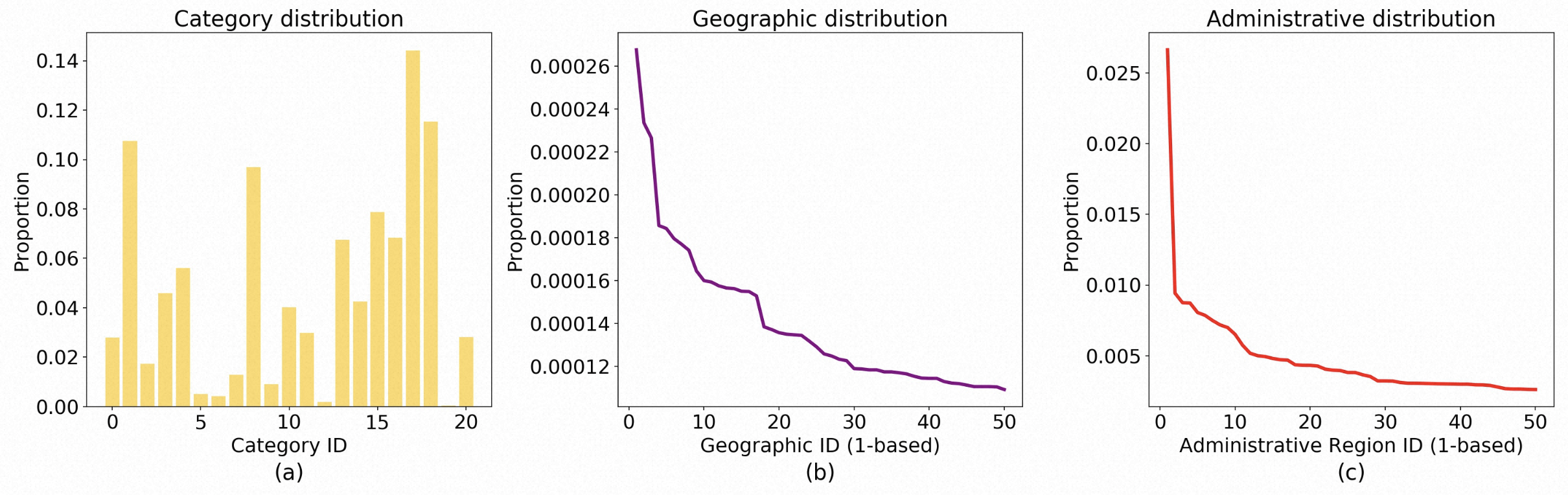}
    \caption{POI distribution across categorical, administrative, and geographic dimensions. For readability, (b) and (c) display only the 50 GIDs and ARIDs with the most POIs. Both the geographic and administrative distributions exhibit a long-tail pattern.}
    \label{Figure_POI_distribution}
\end{figure*}

Figure~\ref{Figure_POI_distribution} presents the POI distributions across category, administrative region, and geographic block.

\subsection{User Profiles}

\begin{figure}[htbp]
    \centering
    \includegraphics[width=0.4\columnwidth]{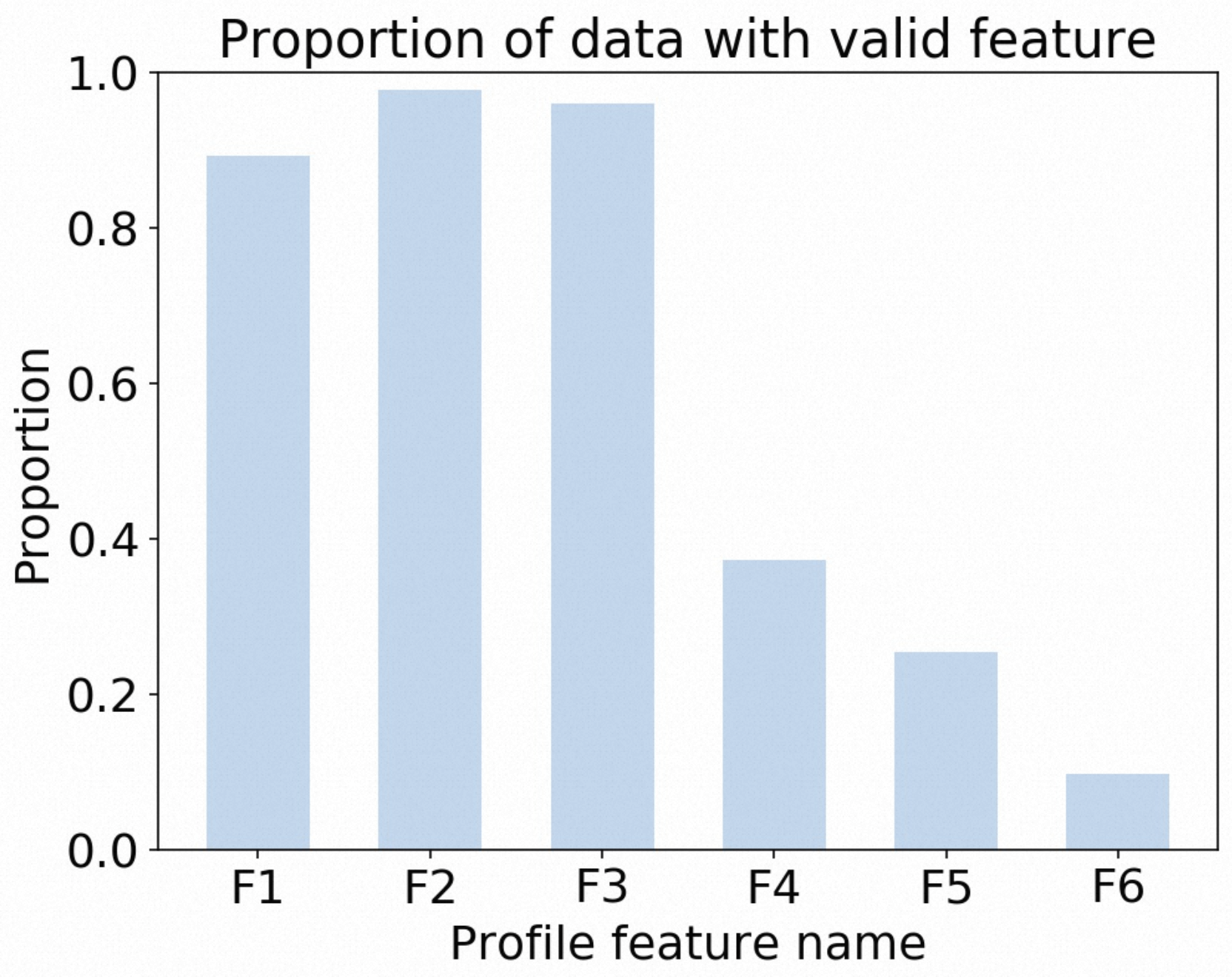}
    \caption{Proportion of valid values for each feature in the IntTravel user-profile data. ``F1'' is short for ``Feature 1.''}
    \label{Figure_user_distribution}
\end{figure}

The IntTravel dataset contains 162,815,693 users, each described by the following fields. Table~\ref{Table_user_examples} provides examples of the user-profile data.

\begin{itemize}
    \item \textbf{User ID}: A unique identifier assigned to each user. To ensure anonymity, these IDs were generated by shuffling the original identifiers and re-indexing them from 0.
    \item \textbf{User Profile Features}: Six distinct profile features. To protect user privacy, both the original feature names and their raw values have been anonymized. Each feature has been transformed into a discrete integer ID, and some values may be missing. Figure~\ref{Figure_user_distribution} shows the proportion of valid, non-missing values for each feature.
\end{itemize}

\begin{table}[htbp]
    \centering
    \caption{Examples of user data, including user IDs and anonymized profile features. ``F1'' is short for ``Feature 1.'' The slash (/) denotes a missing value.}
    \label{Table_user_examples}
    \begin{tabular}{lcccccc}
        \toprule
        \textbf{User ID} & \textbf{F1} & \textbf{F2} & \textbf{F3} & \textbf{F4} & \textbf{F5} & \textbf{F6} \\
        \midrule
        108452457 & 0 & 0 & 2 & / & 2 & 0 \\
        108452458 & 0 & 0 & 6 & 5 & 3 & / \\
        108452459 & 1 & 1 & 1 & / & / & / \\
        \bottomrule
    \end{tabular}
\end{table}

\subsection{User Interaction Behaviors}

\begin{figure*}[htbp]
    \centering
    \includegraphics[width=0.95\linewidth]{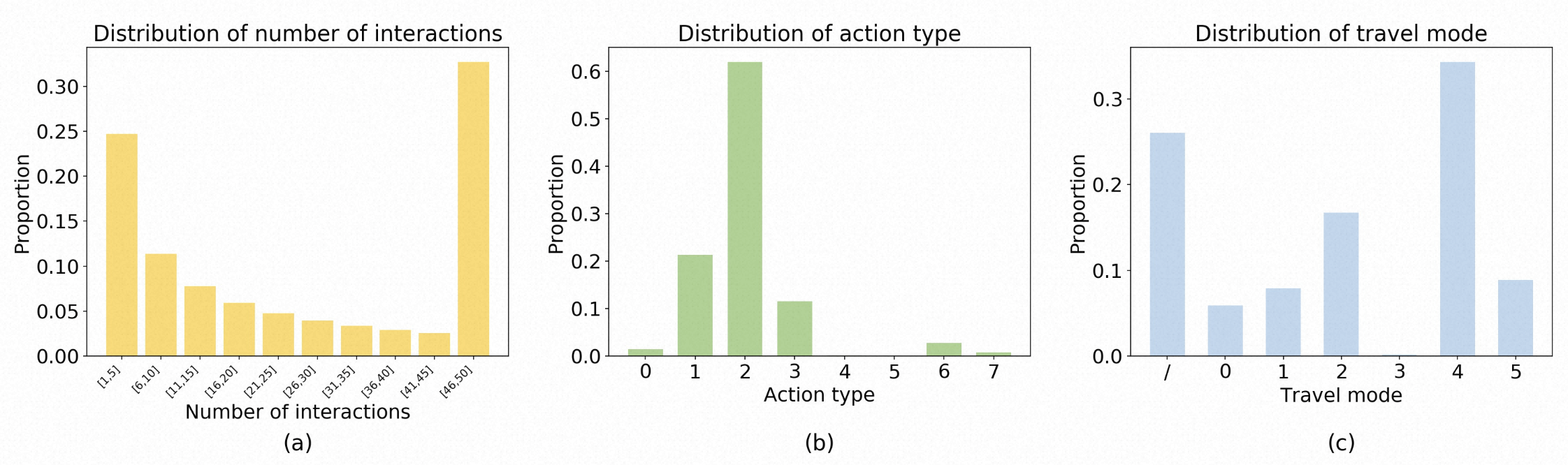}
    \caption{Distribution of user interactions by frequency, action type, and observed travel-mode category. The slash (/) indicates that no travel-mode information is included in a record.}
    \label{Figure_interaction_distribution}
\end{figure*}

The IntTravel dataset includes 4,129,827,011 user interaction events. Each event is characterized by the following fields, with examples provided in Table~\ref{Table_interaction_examples}.

\begin{itemize}
    \item \textbf{User ID}: The identifier of the user who performed the interaction. This ID corresponds to the ``User ID'' in the user dataset.
    \item \textbf{Timestamp}: A value in milliseconds indicating when the interaction occurred. To protect privacy, these values have been transformed and are not the raw timestamps, but their relative order within each user sequence is preserved.
    \item \textbf{Action Type}: An integer ID representing the type of user behavior, such as a click or route planning. The original action names have been anonymized and converted into numerical IDs.
    \item \textbf{Target POI ID}: The identifier of the POI involved in the interaction. This ID corresponds to an entry in the POI dataset.
    \item \textbf{Geographic ID (GID)}: The identifier of the geographic block where the user was located when the interaction occurred. It uses the same encoding system as the GIDs in the POI dataset.
    \item \textbf{Administrative Region ID (ARID)}: The identifier of the administrative region where the user was located when the interaction occurred. It uses the same encoding system as the ARIDs in the POI dataset.
    \item \textbf{Weather}: An integer ID representing the weather condition at the time of the interaction. The original descriptions have been converted into numerical IDs.
    \item \textbf{Travel Mode}: An integer ID representing the user's chosen mode of travel. The original information has been anonymized, and this field may be missing for some records.
    \item \textbf{Via POI ID}: The identifier of a via POI added by the user. It corresponds to an entry in the POI dataset and may be missing because not all interactions involve a via POI.
\end{itemize}

In the released dataset, each historical user-interaction sequence is capped at 50 interactions. This release-specific cap is distinct from the model input lengths used for offline training and online serving. Figure~\ref{Figure_interaction_distribution} presents the interaction distributions along several dimensions.

\begin{table*}[htbp]
    \centering
    \caption{Examples of user interaction log data. Each row is an illustrative interaction record; rows are not presented as a chronological sequence. The slash (/) denotes a missing value.}
    \label{Table_interaction_examples}
    \resizebox{\textwidth}{!}{%
    \begin{tabular}{ccccccccc}
        \toprule
        \textbf{User ID} & \textbf{Timestamp} & \textbf{Action Type} & \textbf{Target POI ID} & \textbf{GID} & \textbf{ARID} & \textbf{Weather} & \textbf{Travel Mode} & \textbf{Via POI ID} \\
        \midrule
        70588810 & 4731399094 & 2 & 1233656 & 24276681 & 2778 & 14 & 1 & / \\
        70588810 & 155476000 & 3 & 908725 & 35758646 & 213 & 9 & 2 & / \\
        77276360 & 6288157866 & 7 & 5852976 & 46698773 & 2883 & 20 & 4 & 5403246 \\
        \bottomrule
    \end{tabular}
    }
\end{table*}

\section{Evaluation Metrics}
\label{Appendix_evaluation_metrics}
Evaluation metrics include both positive indicators (where higher is better) and negative indicators (where lower is better) to provide a holistic view of the model’s capabilities.

\subsection{``When'' Task}

\begin{itemize}
    \item Positive: \textbf{Accuracy (Acc)} measures the percentage of predictions that exactly match the ground-truth departure time, where $\mathcal{T}$ is the test set, $y_i$ is the ground truth, $\hat{y}_i$ is the predicted result, and $\mathbb{I}(\cdot)$ is the indicator function.
    \begin{equation}
        \text{Acc} = \frac{1}{|\mathcal{T}|} \sum_{i \in \mathcal{T}} \mathbb{I}(\hat{y}_i = y_i).
    \end{equation}
    
    \item Negative: \textbf{Mean Absolute Error (MAE)} calculates the average absolute difference between predicted and actual values, reflecting the overall prediction bias.
    \begin{equation}
        \text{MAE} = \frac{1}{|\mathcal{T}|} \sum_{i \in \mathcal{T}} |\hat{y}_i - y_i|.
    \end{equation}
\end{itemize}

\subsection{``How'' Task}

\begin{itemize}
    \item Positive: \textbf{Accuracy (Acc)} is the proportion of correctly predicted travel modes.
    \item Negative: \textbf{Bad Case Rate (BCR)} measures the rate of predictions leading to bad user experience. Mobile navigation applications typically display only three travel mode recommendations on the primary screen. Failing to include the user's preferred option in this top-3 set forces them to take extra steps, resulting in a poor experience. Therefore, we calculate BCR as the \textbf{Top-3 Miss Rate}.
    
\end{itemize}

\subsection{``Where'' and ``Via'' Tasks}

\begin{itemize}
    \item Positive: \textbf{HitRate@N} measures whether the ground-truth POI is included in the top-N recommended items. 
    \begin{equation}
        \text{HitRate@N} = \frac{1}{|\mathcal{T}|} \sum_{i \in \mathcal{T}} \mathbb{I}(y_i \in \text{TopN}(\hat{R}_i)),
    \end{equation}
    where $\hat{R}_i$ is the ranked list of candidate POIs for the $i$-th sample, and $\text{TopN}(\cdot)$ returns the top-N items from that list.
    
    \item Negative: \textbf{Category Inconsistency Rate (CIR)}. Since recommending a POI from the same category as the ground truth can also be reasonable, CIR measures the rate at which the top-1 prediction neither matches the ground-truth POI nor belongs to the same category as the ground truth:
    \begin{equation}
        \text{CIR} = \frac{1}{|\mathcal{T}|} \sum_{i \in \mathcal{T}} \mathbb{I}(\hat{y}_i^{(1)} \neq y_i \land \text{Cate} (\hat{y}_i^{(1)}) \neq \text{Cate} (y_i)),
    \end{equation}
    where $\hat{y}_i^{(1)}$ is the top-1 predicted POI, and $\text{Cate} (\text{POI})$ denotes the category of a given POI.
\end{itemize}

\end{document}